\documentclass[useAMS,usenatbib]{mn2e}
\usepackage{graphicx}

\title[Lithium abundance in a sample of solar-like stars]{Lithium abundance in a sample of solar-like stars}
\author[L{\'o}pez-Valdivia et al.]{R. L{\'o}pez-Valdivia$^1$\thanks{E-mail:
valdivia@inaoep.mx}, J. B. Hern{\'a}ndez-{\'A}guila$^1$,  E. Bertone$^1$, M. Ch{\'a}vez$^1$, \newauthor F. Cruz-Saenz de Miera$^1$, and E. M. Amazo-G{\'o}mez$^2$\\
$^1$Instituto Nacional de Astrof{\'i}sica, {\'O}ptica y Electr{\'o}nica, Luis Enrique Erro 1, Tonantzintla, Puebla, 72840, M{\'e}xico\\
$^2$ Universit{\"a}ts-Sternwarte M{\"u}nchen, Scheinerstrasse 1, D-81679 M{\"u}nchen, Germany}
\begin{document}

\date{Accepted 1988 December 15. Received 1988 December 14; in original form 1988 October 11}

\pagerange{\pageref{firstpage}--\pageref{lastpage}} \pubyear{2002}

\maketitle

\label{firstpage}

\begin{abstract}
We report on the determination of the lithium abundance [A(Li)] of 52
solar-like stars. For 41 objects the A(Li) here presented corresponds
to the first measurement. We have measured the equivalent widths
of the 6708~\AA\ lithium feature in high-resolution spectroscopic images
($R \sim 80\,000$), obtained at the Observatorio Astrof\'isico Guillermo
Haro (Sonora, Mexico), as part of the first scientific observations
of the revitalized Lunar and Planetary Laboratory (LPL) Echelle
Spectrograph, now known as the Cananea High-resolution Spectrograph
(CanHiS).  Lithium abundances were derived with the Fortran code MOOG,
using as fundamental input a set of atmospheric parameters recently obtained
by our group.\\
With the help of an additional small sample with previous A(Li)
determinations, we demonstrate that our lithium abundances are in agreement,
to within uncertainties, with other works. Two target objects stand out from
the rest of the sample. The star BD+47~3218 ($T_{\rm eff}$~=~6050$\pm$52~K,
A(Li)~=~1.86$\pm$~0.07~dex) lies inside the so-called {\em lithium desert} in
the the A(Li)--$T_{\rm eff}$ plane.  The other object, BD+28~4515, has
an A(Li)~=~3.05$\pm$0.07~dex, which is the highest of our sample and
compatible with the expected abundances of relatively young stars.
\end{abstract}


\section{Introduction}
In convective stars, the photospheric lithium abundance A(Li)\footnote{A(Li) = $\log$($n$(Li)/$n$(H)) + 12, where $n$ is the number 
density of atoms.} decreases with time, because vertical motions transport Li atoms to deeper layers hot enough ($\approx 2.5\times10^{6}$~K) to 
burn it by proton capture.

The Sun has a very low photospheric A(Li) compared to nearby solar analogs \citep[e.~g.,][]{lambert04}, and it is more than 2~dex lower than the 
meteoric abundance (A(Li)$_\odot = 1.05 \pm 0.10$ vs. A(Li)$_{\rm{met}} = 3.26 \pm 0.05$; \cite{asplund09}).
This difference cannot be explained by 
standard models \cite[e.~g.,][]{dantona94}, consequently, several  processes  have been proposed: for instance, mass 
loss \citep{swenson92}, angular momentum loss and rotation \citep{schatzam91,pinsonneault92,deliyanis97,eggenberger10}, magnetic fields
\citep{eggenberger10,li14} or gravity waves 
\citep{garcia91,montalban00,talon05}, all of which induce an extra mixing.\\
More recently, it has also been suggested that the presence of planets could affect the evolution of Li abundance.
\cite{bouvier08} and \cite{eggenberger12} proposed that a long lasting star-disk interaction during the pre-main-sequence stage could slow down 
the rotation of the host star, enhancing the rotational mixing and thus destroying more lithium. 
\cite{theado12} investigated the effects of disk matter accretion, that induce diffusive instabilities resulting in an extra mixing episode.    
\cite{castro09}  pointed out that a transfer of 
angular momentum due to planetary migration could increase the mixing efficiency and Li depletion. Additionally, \cite{takeda10} confirmed that 
Li depletion in solar-like stars is more efficient when stellar rotation 
decelerates.\\ 
Even though some authors \citep[e.~g.,][]{israelian04,gonzalez08,israelian09,gonzalez14,delgado14,figueira14,gonzalez15} claimed to have found 
that stars with exoplanets show an excess of Li depletion, the existence of a possible A(Li)-planet correlation is still matter of debate, as  
different works \citep{ryan00,luck06,ghezzi10,baumann10,ramirez12} had not found such correlation.
Increasing the number of main-sequence (MS) stars with known Li abundance is therefore helpful to reach more robust statistical results.

In this work, we present the determination of lithium abundances, measured from the 6708~\AA\ resonance line, for 52 solar-like stars of intermediate brightness. For 41 of the targets
our analysis represents the first Li measurement. The working sample and observations are described in Section 2. In Section 3, we detail the procedure followed for deriving abundances and, in Section 4, we discuss the results.

\section{Sample Definition and Observations}

The stellar sample is somewhat inhomogeneous, but reflects a continuation of 
a global project aimed at determining stellar parameters 
(T$_{\rm eff}$ / $\log g$ / \rm{[M/H]}) as well as abundances of individual 
chemical species of a sample of Sun-like stars (types G0-G3) on 
the MS. As an initial step, we have selected 30 stars from our previous 
work \citep{lopezval14}, 12 objects of which correspond to 
metal-rich stars (SMR; \rm{[M/H]}$\geq$0.16~dex). This latter sub-sample 
is in fact composed by all the SMR stars that are sufficiently bright 
to allow high 
resolution observations with the instrument briefly described below. The 
sample is complemented with 11 stars, that accomplish the criteria in 
spectral type and luminosity class of \cite{lopezval14}, included in the 
catalog of \cite{casagrande11} for which no previous Li
determination has been reported. For comparison purposes, we also 
observed 11 stars from the work of \cite{ramirez12} who provide their Li
abundances.

The full sample is listed in Table~\ref{tab:samp}, where we give some 
basic data and the results of our analysis. Columns 1-13 provide the star 
identification, the spectral type, the leading atmospheric parameters and their uncertainties \citep[from][]{lopezval14,casagrande11},
the 6708~\AA\ Li line equivalent width (EW) and its uncertainty, as described in 
the following section, and the last column gives a label for the source 
of stellar parameters.

\begin{figure*}
\includegraphics[width=145mm]{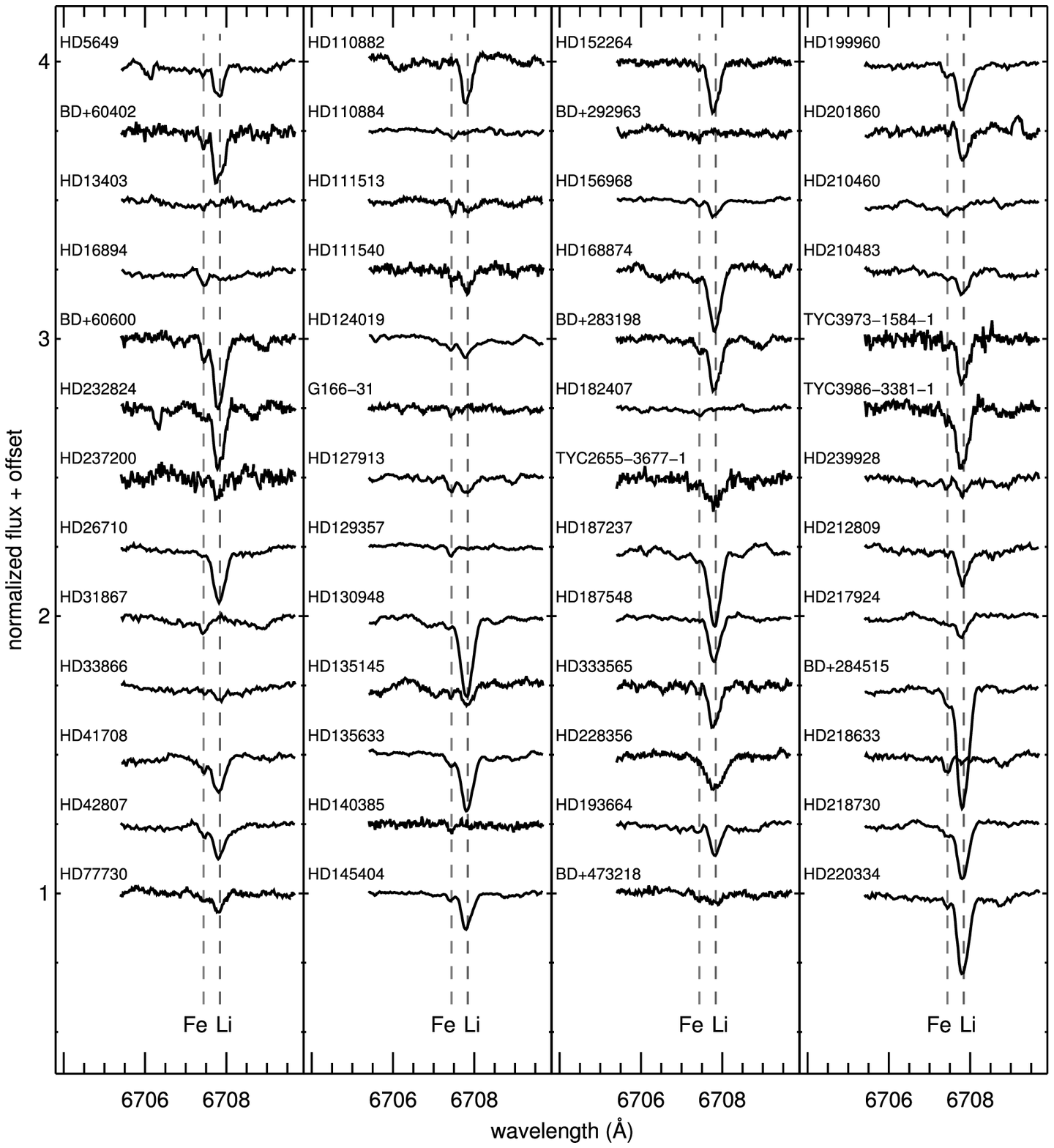}
\caption{Normalized spectra of our stellar sample, centered around the Li line. The position of the Li line and of the adjacent Fe line is indicated by vertical dashed lines.}
\label{fig:obj}
\end{figure*}

The spectroscopic data were collected at the 2.1~meter telescope of the Observatorio Astrof\'isico Guillermo Haro, located in Mexico, using the Cananea High-resolution Spectrograph (CanHiS), as part of the first scientific observations of this instrument\footnote{CanHiS was donated in 2009 by the University of Arizona to our institute; it was originally known as LPL Echelle Spectrograph 
and used at the 61-inch Catalina Telescope to carry out observations of Solar System bodies at specific spectral regions \citep[for more details about 
its original design see][]{hunten91}.}, at this facility. CanHiS is equipped with filters, that provide access to  $\sim$40~\AA-wide wavelength intervals in single 
diffraction orders. 
Our observations were centered at 6710~\AA, with a spectral resolving power of $R\sim 80000$ and a typical signal-to-noise ratio (S/N) of about 100. All the data collected (at least 3 images for each star) were reduced with IRAF, following standard procedures: 
bias substraction, flat-field correction, cosmic-ray removal, wavelength calibration through an internal UNe lamp and, finally, continuum normalization. 
We then transformed all the spectra to the rest frame, using the spetrum of the Sun by \cite{kurucz84} as template. 
For each star we co-added its spectra weighted by their 
S/N. In Fig.~\ref{fig:obj} we present the spectra of the Li region of the 52 solar-like stars observed in this work.

\begin{table*}
\begin{minipage}{180mm}
\caption{Physical parameters and Li abundance of the stellar sample. Temperatures from \citet{casagrande11} were corrected following Eq.~\ref{eq:cal}.}
\label{tab:samp}
\begin{tabular}{lrlrrrrrrrrrrc}
\hline
Object  &   V   &  SType & $T_{\rm eff}$  & $\sigma$ & $\log g$ & $\sigma$  & $\rm{[M/H]}$ & $\sigma$  & EW(Li)& $\sigma$EW(Li) & A(Li) & $\sigma$A(Li) & S \\
        &       &       &  (K)   & (K) & (dex) & (dex) & (dex) & (dex) & (m\AA) & (m\AA) & (dex) & (dex)& \\
\hline  
HD 5649	                                                                &	8.70	&	G0V	&	5830	&	52	&	4.45	&   0.22  &  -0.08  &   0.04  &   27.5	&	3.6	&	2.05	&	0.08	&	L	\\
BD+60 402	                                                        &      10.26    &	G0V	&	5985	&	72	&	4.30	&   0.40  &   0.22  &   0.09  &   58.0	&	9.5	&	2.58	&	0.12	&	L	\\
HD 13403                                                                &	7.00	&	G3V	&	5565	&	204	&	4.74	&    -    &  -0.29  &    -    &   $<$ 5.5	&	-	&	$<$ 1.04	&	-	&	C	\\
HD 16894    	                                                        &	8.02	&	G2V	&	5500	&	70	&	4.05	&   0.30  &  -0.10  &   0.09  &   6.4&	1.0&	1.1	&	0.09	&	L	\\
BD+60 600	                                                        &	8.65	&	G0V	&	5655	&	47	&	3.95	&   0.20  &   0.20  &   0.07  &   78.0	&	1.2	&	2.51	&	0.04	&	L	\\
HD 232824	                                                        &	9.52	&	G2V	&	5900	&	67	&	4.15	&   0.35  &   0.16  &   0.08  &   61.7	&	5.7	&	2.56	&	0.08	&	L	\\
HD 237200	                                                        &	9.66	&	G0V	&	6045	&	55	&	4.25	&   0.32  &   0.18  &   0.05  &   18.3	&	3.1	&	2.03	&	0.09	&	L	\\
HD 26710	                                                        &	7.18	&	G2V	&	5815	&	47	&	4.55	&   0.20  &  -0.04  &   0.04  &   60.8	&	7.1	&	2.47	&	0.08	&	L	\\
HD 31867	                                                        &	8.05	&	G2V	&	5590	&	57	&	4.40	&   0.25  &  -0.10  &   0.06  &   $<$ 4.9	&	-	&	$<$ 1.03	&	0.08	&	L	\\
HD 33866	                                                        &	7.87	&	G2V	&	5481	&	123	&	4.33	&    -    &  -0.07  &    -    &   11.0	&	1.6	&	1.28&	0.13	&	C	\\
HD 41708	                                                        &	8.03	&	G0V	&	5998	&	58	&	4.55	&    -    &   0.08  &    -    &   38.1	&	1.2	&	2.36&	0.05	&	C	\\
HD 42807\footnote{Star with previous determination of A(Li).\label{fn}} &	6.44	&	G2V	&	5617    &	80	&	4.53	&    -    &  -0.11  &    -    &   43.8	&	5.5	&	2.10	&	0.10	&	C	\\
HD 77730	                                                        &	7.39	&	G2V	&	5698    &	80	&	4.13	&    -    &  -0.05  &    -    &   18.9	&	2.8	&	1.75&	0.10	&	C	\\
HD 110882	                                                        &	8.87	&	G1V	&	5880	&	50	&	4.40	&   0.25  &  -0.28  &   0.04  &   38.8	&	0.8	&	2.26	&	0.05	&	L	\\
HD 110884	                                                        &	9.11	&	G3V	&	5905	&	87	&	4.30	&   0.40  &  -0.26  &   0.08  &   $<$ 3.9	& - &	$<$ 1.20	&	-	&	L	\\
HD 111513\textsuperscript{\ref{fn}}	                                &	7.35	&	G1V	&	5723    &	80	&	4.31	&    -    &   0.12  &    -    &   11.4	&	1.7	&	1.53	&	0.10	&	C	\\
HD 111540	                                                        &	9.54	&	G1V	&	5840	&	47	&	4.20	&   0.25  &   0.14  &   0.05  &   23.1	&	5.2	&	1.97	&	0.12	&	L	\\
HD 124019	                                                        &	8.56	&	G2V	&	5685	&	57	&	4.65	&   0.25  &  -0.18  &   0.06  &   19.0	&	2.6	&	1.73	&	0.08	&	L	\\
HD 126991	                                                        &	7.90	&	G2V	&	5360	&	107	&	3.15	&   0.40  &  -0.34  &   0.14  &   $<$ 4.7	&	-	&	$<$ 0.79	&	-	&	L	\\
HD 127913	                                                        &	9.14	&	G2V	&	5475	&	65	&	4.45	&   0.20  &  -0.12  &   0.08  &   17.5	&	1.2	&	1.50	&	0.07	&	L	\\
HD 129357	                                                        &	7.83	&	G2V	&	5775	&	52	&	4.30	&   0.22  &  -0.14  &   0.05  &   $<$ 4.4	&	-	&	$<$ 1.14	&	-	&	L	\\
HD 130948\textsuperscript{\ref{fn}}	                                &	5.88	&    F9IV-V     &	5885    &	80	&	4.42	&    -    &  -0.09  &    -    &   93.8	&	10.0	&	2.84	&	0.12	&	C	\\
HD 135145\textsuperscript{\ref{fn}}	                                &	8.35	&	G0V	&	5997	&	80	&	4.14	&    -    &  -0.02  &    -    &   20.3	&	2.8	&	2.04&	0.10	&	C	\\
HD 135633	                                                        &	8.46	&	G0V	&	6095	&	67	&	4.25	&   0.40  &   0.22  &   0.06  &   69.0	&	0.9	&	2.79	&	0.06	&	L	\\
HD 140385\textsuperscript{\ref{fn}}	                                &	8.57	&	G2V	&	5735	&	60	&	4.60	&   0.27  &  -0.16  &   0.08  &   $<$ 5.6	&	-	&	$<$ 1.21	& -	&	L	\\
HD 145404	                                                        &	8.54	&	G0V	&	5920	&	82	&	4.43	&    -    &  -0.16  &    -    &   37.2	&	4.6	&	2.28&	0.10	&	C	\\
HD 152264	                                                        &	7.74	&	G0V	&	6177	&	73	&	4.09	&    -    &   0.02  &    -    &   53.6	&	6.9	&	2.70&	0.10	&	C	\\
BD+29 2963	                                                        &	8.42	&	G0V	&	5865	&	55	&	4.70	&   0.22  &   0.00  &   0.04  &   $<$ 5.9	&	-	&	$<$ 1.36	&	-	&	L	\\
HD 156968\textsuperscript{\ref{fn}}	                                &	7.97	&	G0V	&	6105	&	96	&	4.42	&    -    &  -0.03  &    -    &   18.9	&	1.4	&	2.09&	0.09	&	C	\\
HD 168874\textsuperscript{\ref{fn}}	                                &	7.01	&	G2IV	&	5696    &	80	&	4.41	&    -    &  -0.05  &    -    &   65.4	&	9.6	&	2.41	&	0.12	&	C	\\
BD+28 3198	                                                        &	8.66	&	G2V	&	5840	&	35	&	4.00	&   0.17  &   0.24  &   0.05  &   59.1	&	8.1	&	2.49	&	0.09	&	L	\\
HD 182407	                                                        &	7.77	&	G0V	&	5953    &	70	&	3.94	&    -    &   0.25  &    -    &   $<$ 3.6	&	-	&	$<$ 1.22&	-	&	C	\\
TYC 2655-3677-1	                                                        &       9.93    &	G0V     &	6220	&	47	&       4.15	&   0.27  &   0.28  &   0.05  &   38.9	&	6.2	&	2.54	&	0.10	&	L	\\
HD 187237\textsuperscript{\ref{fn}}	                                &	6.88	&    G2IV-V     &       5687    &	62	&	4.42	&    -    &  -0.04  &    -    &   89.2	&	9.8	&	2.63	&	0.10	&	C	\\
HD 187548	                                                        &	7.98	&	G0V	&	6170	&	42	&	4.80    &   0.17  &  -0.24  &   0.03  &   49.3	&	2.4	&	2.63	&	0.05	&	L	\\
HD 333565	                                                        &	8.75	&	G1V	&	5990	&	52	&	4.45	&   0.27  &   0.12  &   0.05  &   44.2	&	6.5	&	2.43	&	0.09	&	L	\\
HD 228356	                                                        &	9.07	&	G0V	&	6055	&	37	&	4.00    &   0.20  &   0.16  &   0.05  &   63.4	&	5.6	&	2.71	&	0.07	&	L	\\
HD 193664\textsuperscript{\ref{fn}}	                                &	5.93	&	G3V	&	5942    &	112	&	4.47	&    -    &  -0.11  &    -    &   30.2	&	2.6	&	2.19&	0.11	&	C	\\
BD+47 3218	                                                        &	8.70	&	G0V	&	6050	&	52	&	4.05	&   0.30  &   0.16  &   0.06  &   12.6&	1.30	&	1.86	&	0.07	&	L	\\
HD 199960\textsuperscript{\ref{fn}}	                                &	6.21	&	G1V	&	5895    &	80	&	4.24	&    -    &   0.22  &    -    &   61.0	&	8.1	&	2.54&	0.11	&	C	\\
HD 201860	                                                        &	8.65	&	G0V	&	5620    &	107	&	3.66	&    -    &  -0.02  &    -    &   30.9	&	0.3	&	1.93&	0.10	&	C	\\
HD 210460   	                                                        &	6.19	&	G0V	&	5357    &	80	&	3.58	&    -    &  -0.17  &    -    &   $<$ 4.8	&	-	&	$<$ 0.80	&	-	&	C	\\
HD 210483\textsuperscript{\ref{fn}}	                                &	7.59	&	G1V	&	5878    &	95	&	4.19	&    -    &  -0.01  &    -    &   19.8	&	2.7	&	1.93&	0.11	&	C	\\
TYC 3973-1584-1                                                         &      10.74    &	G2V	&	6000	&	45	&	4.45	&   0.22  &   0.20  &   0.05  &   47.2	&	6.4	&	2.48	&	0.09	&	L	\\
TYC 3986-3381-1	                                                        &      10.37    &	G2V	&	5855	&	57	&	4.15	&   0.25  &   0.26  &   0.07  &   79.1	&	5.1	&	2.68	&	0.07	&	L	\\
HD 239928	                                                        &	8.69	&	G2V	&	5870	&	67	&	4.40	&   0.30  &   0.06  &   0.06  &   17.3	&	2.4	&	1.86	&	0.09	&	L	\\
HD 212809	                                                        &	8.64	&	G2V	&	5975	&	55	&	4.55	&   0.27  &   0.16  &   0.05  &   33.5	&	2.6	&	2.27	&	0.06	&	L	\\
HD 217924	                                                        &	7.22	&	G0V	&	5900	&	52	&	4.85	&   0.20  &  -0.10  &   0.05  &   17.2	&	3.6	&	1.87	&	0.11	&	L	\\
BD+28 4515	                                                        &	8.73	&	G2V	&	5580	&	40	&	3.50	&   0.17  &  -0.22  &   0.06  &   146.0	&	6.8	&	3.05	&	0.07	&	L	\\
HD 218633	                                                        &	8.19	&	G2V	&	5385	&	70	&	3.50	&   0.30  &  -0.32  &   0.10  &   $<$ 6.4	&	-&	$<$ 0.96	& -	&	L	\\
HD 218730	                                                        &	7.32	&	G0V	&	5922	&	83	&	4.45	&    -    &   0.09  &    -    &   54.8	&	9.5	&	2.50	&	0.12	&	C	\\
HD 220334	                                                        &	6.62	&	G2V	&	5801    &	91	&	4.13	&    -    &   0.14  &    -    &   87.9	&	3.0	&	2.72	&	0.09	&	C	\\\hline
\end{tabular}
\end{minipage}
\end{table*}

\section{Abundances Determination}
In this Section we will present the procedure that we carried out for measuring the Li abundance of the stellar sample.

\subsection{Photospheric parameters and model atmospheres}
\label{model}
We collected the $T_{\rm eff}$,  the surface gravity ($\log{g}$), the global metallicity ([M/H]) and their uncertainties (see Table~\ref{tab:samp}, column 4 to 9) from two works.
The first one is \cite{lopezval14}, where $T_{\rm eff}$, $\log{g}$ and ${\rm[M/H]}$ were simultaneously determined using a set of Lick-like
indices and intermediate-resolution spectroscopic observations.
The second work is  \cite{casagrande11}, where the $T_{\rm eff}$ is obtained through the infrared flux method \citep{blackwell77,casagrande06}, 
the $\log{g}$ from the fundamental relation, which involves bolometric luminosity and $T_{\rm eff}$ and finally the metallicity as ${\rm[Fe/H]}$ from a 
calibration of Str{\"o}mgren colours\footnote{We used the overall metallicity reported in the electronic version of the catalog of 
\cite{casagrande11}, instead of ${\rm[Fe/H]}$.}.
The last column of Table~\ref{tab:samp} reports the source of the stellar parameters for each star of our sample.
 
\begin{figure}
\includegraphics[width=77mm]{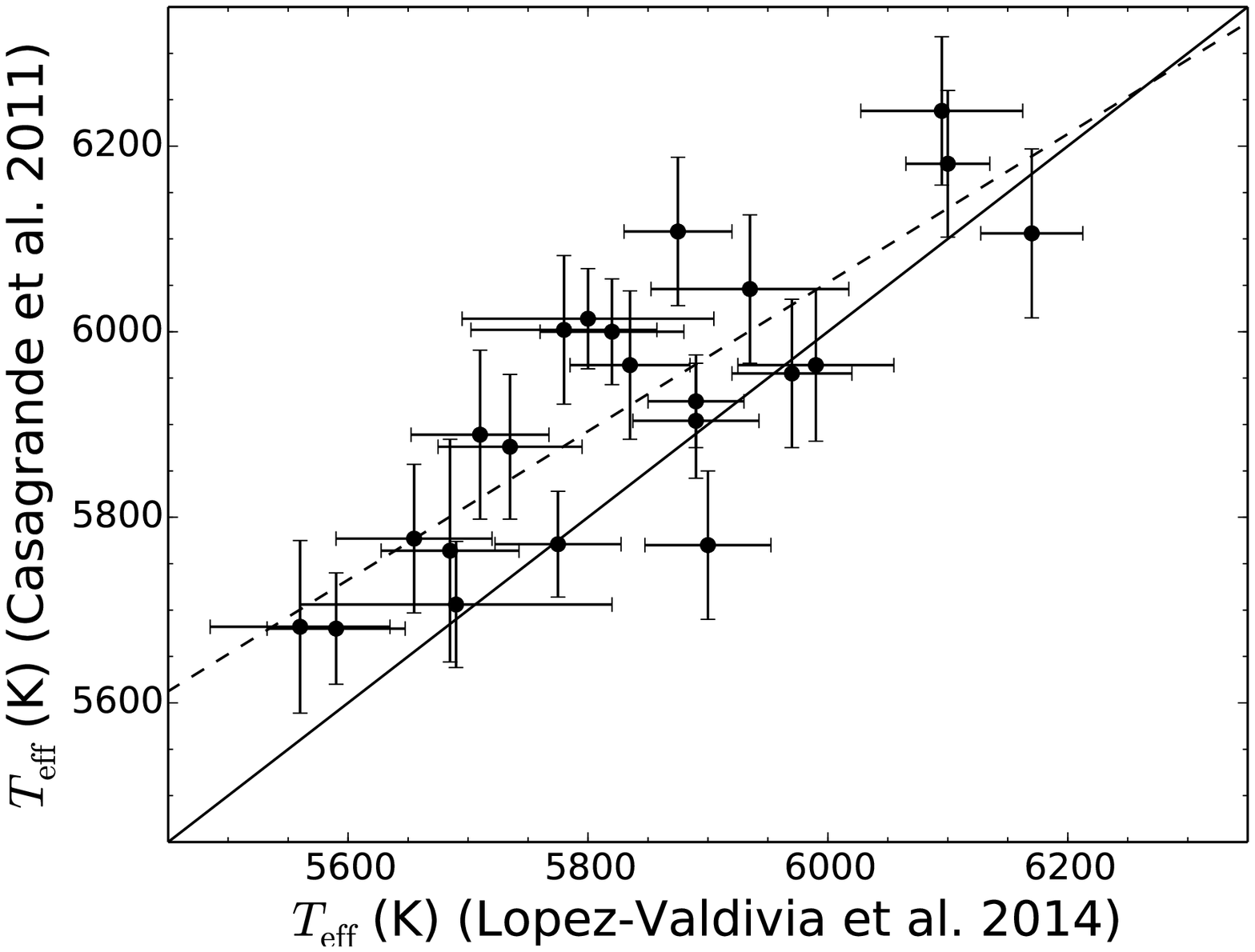}
\caption{Comparison of $T_{\rm eff}$ from \citet{lopezval14} and \citet{casagrande11} for 22 solar-like stars in common. The dashed line shows the best linear regression, while the solid line indicates the one-to-one relation.}
\label{fig:cal}
\end{figure}

By means of the 22 object in common in the \cite{lopezval14} and \cite{casagrande11} samples, we investigated possible systematic differences in the $T_{\rm eff}$ scales, which is the most important parameter for Li abundance determination (see Sect.~\ref{sec:err}). The data, plotted in Fig.~\ref{fig:cal}, indeed show a tendency of \cite{casagrande11} temperatures to be slightly higher, mostly at the cooler end.
Since \cite{lopezval14} $T_{\rm eff}$ scale was constructed to optimize the parameters of G0-G3 MS stars and considering that \cite{casagrande11} stated that their $T_{\rm eff}$ are on average 100~K hotter that previous analyses, we transformed the \cite{casagrande11} temperatures to the \cite{lopezval14} scale through a linear regression of data displayed in Fig.~\ref{fig:cal}:
\begin{equation}
T_{\rm eff,cor} = \frac{T_{\rm eff, Cas} - 1247.86}{0.80}
\label{eq:cal}
\end{equation}  
where $T_{\rm eff, Cas}$ is the temperature reported in \cite{casagrande11} and  $T_{\rm eff,cor}$ is the corrected one.

We then determined a model atmosphere for each star by interpolating  the \cite{castelli03} grid of models in the $T_{\rm eff}$, $\log{g}$ and 
${\rm[M/H]}$ 3-dimensional space.
For all cases, we used the microturbulence velocity $\xi = 2$~km~s$^{-1}$ (see also Section~\ref{sec:err}).

\subsection{Equivalent widths}

We measured the EW of the 6708~\AA\ lithium line with a flux fitting procedure. We simultaneously fit the Li and the adjacent Fe~\textsc{i} line at 6707.44~\AA\ using three Gaussian functions: two of them are necessary to reproduce the two-component (simply called Li$_1$ and Li$_2$) asymmetric Li line, while the third is needed to quantify the Fe contamination.
The parameters of the Gaussian functions (central wavelength, width and amplitude) were modified within appropriate values for each star to obtain the best fit, which was also validated through visual inspection.
The integral of the sum of the Li$_1$ and Li$_2$ Gaussians provides the EW(Li).

We adopted a Monte Carlo procedure to estimate the error in the measurement of EW(Li). For each star,
we determined the standard deviation of the flux in two wavelength regions, 6705.4--6707.0~\AA\ and 6708.6--6709.7~\AA, bracketing the Li line. We used this value as the width of a Gaussian distribution which provides the random noise value that we added to the observed spectrum and we repeated the fitting procedure previously described. We carried out 1000 realizations to obtain the EW distributions of the three Gaussian functions, whose standard deviations we added in quadrature to obtain the error $\sigma$EW(Li).
The EW(Li) values, along with their errors, are reported in columns 10 and 11 of Table~\ref{tab:samp}.

\subsection{Lithium abundances computation}
To carry out the determination of the Li abundances, we made use of the driver {\it abfind} of the February 2013 version of the Fortran code 
MOOG \citep{sneden73}, which performs an adjustment of the abundances to match a single-line equivalent width. This 
code requires a model atmosphere and the physical parameters of a list of absorption lines to compute atomic abundances.
We assumed the Li line to be a single electronic transition centered at $\lambda$6707.84~\AA\ with $\log{gf} = +0.167$, that corresponds to the average wavelength and the sum of the $gf$ values of the two components 
(Li$_1$: $\lambda$=6707.76~\AA, $\log{gf}$=$-0.009$; Li$_2$: $\lambda$=6707.91~\AA, $\log{gf}$=$-0.309$), obtained from the Vienna Atomic Line Database \citep[VALD;][]{kupka99}.

\subsubsection{Error of Li abundances}
\label{sec:err}
The main sources of error in the determination of the Li abundance are related with uncertainties on the stellar parameters and on the EW
measurement. In order to have an estimation of the errors introduced by the atmospheric parameters, we conducted some theoretical tests. We varied, one at a time, the four main stellar 
parameters ($T_{\rm eff}$, $\log g$, ${\rm[M/H]}$ and $\xi$), maintaining the other three fixed, and compute the Li abundance, assuming an 
EW(Li)~=~50~m\AA, to measure the variations in A(Li). The chosen fixed values were 5750~K, 4.5~dex, 0.0~dex, and 2.0~km~s$^{-1}$, for $T_{\rm eff}$, $\log g$, ${\rm[M/H]}$ and $\xi$, respectively.
We obtained the following results: a change in effective temperature of $\Delta T_{\rm eff}=100$~K produces a lithium abundance variation of $\Delta {\rm A(Li)} = 0.09$~dex, while a  very small change in abundance of $\Delta {\rm A(Li)} = 0.01$~dex is obtained if we modify the other parameters by $\Delta \log{g}=0.5$~dex, $\Delta {\rm[M/H]}=0.15$~dex and $\Delta \xi = 1$~km~s$^{-1}$.
These results, very similar to those from previous works \citep[e.~g.,][]{chen01,charbonel05}, indicate that the uncertanty in the $T_{\rm eff}$ is by far the dominant source of error. Furthermore, our assumption of a single value for $\xi$ of our models does not introduce a significant systematic error in the Li abundance. This choice is also supported, for instance, by the work of \cite{gray01}, who derived a $\xi$ between 1 and 2~km~s$^{-1}$ for their sample of F7--G2 MS stars.

We calculated the contribution of the EW uncertanty to the error on the Li abundance [$\sigma$A(Li)] by computing the A(Li) for the values EW(Li)$\pm \sigma$EW(Li) and dividing their difference by half. 
The total $\sigma$A(Li) is the sum in quadrature of the errors on  $T_{\rm eff}$ and EW. All results are reported in Table~\ref{tab:samp}.

\begin{table*}
\caption{The lithium abundances and stellar parameters of the stars in common with \citet{ramirez12}.}
\label{tab:comp}
\begin{tabular}{lrcccccc}
\hline
Object  &   A(Li)& $\sigma$A(Li)& $T_{\rm eff}$ & $\log g$ & $\rm{[Fe/H]}$ & A(Li)& $\sigma$A(Li)\\
        &  Our   &   Our      & Ramirez & Ramirez & Ramirez & Ramirez & Ramirez \\

\hline  
HD 42807  & 2.10 & 0.10 & 5674 & 4.51 & -0.04 & 1.92 & 0.09 \\   
HD 111513 & 1.53 & 0.10 & 5811 & 4.32 &  0.11 & 1.44 & 0.03 \\   
HD 130948 & 2.84 & 0.12 & 5942 & 4.39 & -0.11 & 2.86 & 0.07 \\   
HD 135145 & 2.04 & 0.10 & 5858 & 4.09 & -0.08 & 1.80 & 0.04 \\   
HD 156968 & 2.09 & 0.09 & 5901 & 4.36 & -0.14 & 1.92 & 0.02 \\   
HD 168874 & 2.41 & 0.12 & 5858 & 4.44 &  0.01 & 2.68 & 0.10 \\   
HD 187237 & 2.63 & 0.10 & 5792 & 4.43 &  0.02 & 2.18 & 0.10 \\   
HD 193664 & 2.19 & 0.11 & 5902 & 4.42 & -0.14 & 2.22 & 0.02 \\   
HD 199960 & 2.54 & 0.11 & 5833 & 4.20 &  0.13 & 2.43 & 0.10 \\   
HD 210483 & 1.93 & 0.11 & 5834 & 4.14 & -0.12 & 1.97 & 0.15 \\   
\hline
\end{tabular}
\end{table*}

\section{Discussion}

In order to put the Li abundance of our sample in the context of the known results for solar-like stars, we used the work of \citet{ramirez12} as a reference, as it is, to date, the largest homogeneous compilation of lithium abundance of MS and subgiants stars.
They derived stellar parameters and Li abundances for 671 stars and included measurements from the literature to construct a catalog 
of 1381 FGK dwarf and subgiant stars. Analyzing the distribution of data in the A(Li)--$T_{\rm eff}$ plane, they confirmed 
that, in the interval  $5950 \leq T_{\rm eff} \leq 6100$~K, stars appear to be neatly separated into two groups of high and low Li abundance, 
creating an avoidance region, called {\it lithium desert}, as was suggested for the first time by \cite{chen01}.
These authors proposed that the low-Li side of the A(Li)--$T_{\rm eff}$ diagram is populated by evolved lithium dip objects \citep{boesgaard86}, that might burn the Li during the pre-MS stage. However, \citet{ramirez12} argues against this suggestion, based on their analyisis of the mass-metallicity correlation \citep[see also][]{balachandran95,cumings12} of their whole sample, which includes the \citet{chen01} data. They also show that there are no young stars 
(age$<$2~Gyr) at the lower side of the {\em lithium desert}, suggesting that the mechanism that depletes surface lithium in the objects that populate that region should take place during the MS or subgiant phases. However, \citet{ramirez12} do not further speculate on the nature of this mechanism.

\begin{figure}
\includegraphics[width=75mm]{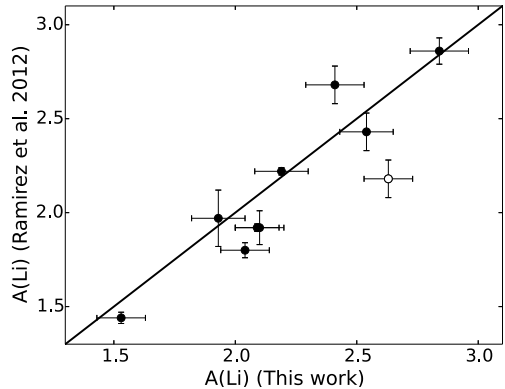}
\caption{Comparison of the lithium abundance of the stars in common with \citet{ramirez12}. HD~187237 (see discussion in text) is represented by an empty square, while the solid line indicates the one-to-one relation.}
\label{fig:comp}
\end{figure}

In total, we have 11 stars in common with \cite{ramirez12}; however, for the star HD~140385 we have only upper limits  and was
excluded  from  the comparison. In Fig.~\ref{fig:comp}, we present the comparison for the remaining common 10 stars, whose A(Li) are reported in Table~\ref{tab:comp}.
The star HD~187237 presents the largest discrepancy of 0.45~dex. In order to achieve the same lithium 
abundance reported in \citet{ramirez12}, this object should have half the EW measured in our work, which would represent a 4.5$\sigma$ anomaly. 
Since, the EW measurement for this 
star is one of our best determinations, and the stellar parameters used in both works are not so different, we do not have a clear cause for the
discrepancy. In the following analysis, we exclude this star too. 
In general, we found a good agreement between the A(Li) determined in this work and those determined in \cite{ramirez12}: on average, we found a difference of +0.05~dex (our minus \citealt{ramirez12} data). This offset is partly caused by adopting different stellar parameters (mainly different $T_{\rm eff}$) and/or different atomic data.

\begin{figure*}
\includegraphics[width=150mm]{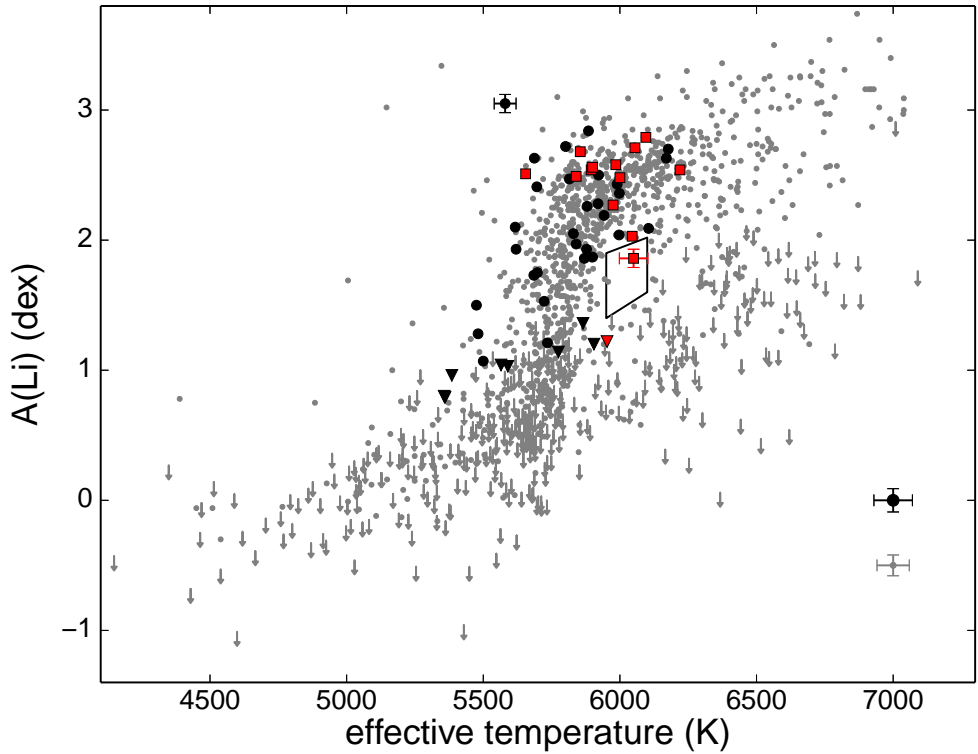}
\caption{Distribution of lithium abundance with $T_{\rm eff}$. Black filled circles indicate stars in our sample with [M/H]$<$0.16~dex, red squares show our SMR stars, and downward triangles mark 3$\sigma$ upper limits. Grey dots and downward arrows represent 
\citet{ramirez12} determinations and upper limits. The polygon shows the so called {\it lithium desert}. At the bottom
right of the panel, we show the average errors for this work (black) and \citet{ramirez12} (gray).}
\label{fig:ram}
\end{figure*}

Figure~\ref{fig:ram} presents the A(Li) vs. $T_{\rm eff}$ distribution of our results and their comparison with the complete \cite{ramirez12} compilation. It allows us to address three issues.

\subsection*{Super metal-rich stars.}
Our stellar sample includes 12 SMR stars of \cite{lopezval14} ([M/H]$>$+0.16~dex) plus 2 from \cite{casagrande11}, indicated with red symbols in Fig.~\ref{fig:ram}.  
It is worthwhile mentioning that SMR stars have become a particularly interesting class, not only because they are important for modeling the chemical evolution of the Galaxy, but also for their positive correlation with the presence of giant planets \citep[e.~g.,][]{gonzalez98,santos01,fischer05} .

The distribution of the Li abundances of our SMR stars is similar to that of the rest of our sample and it also resembles that of the SMR stars in \cite{ramirez12}. All but three objects lie in the high abundance range of $2.25 < {\rm A(Li)} < 2.80$~dex. In the case of the coolest object, BD+60~600, its A(Li)=2.51~dex is quite higher than the median value for its $T_{\rm eff}$ class. On the warmer side, 3 stars show a lower than average Li abundance (A(Li)$<$2.10~dex): HD~237200, BD+47~3218, and HD~182407. The case of BD+47~3218 is discussed below in more detail.

\subsection*{BD+47~3218: the star in the {\em lithium desert}.}

The star BD+47~3218 has $T_{\rm eff}$=6050$\pm$52~K and A(Li)=1.86$\pm$0.07~dex: these values locate it inside the {\it lithium desert}, at 1$\sigma$ in the case of $T_{\rm eff}$ and at 1.7$\sigma$ in the case of Li abundance. If we consider our possible small overestimation of the A(Li) with respect to the \cite{ramirez12} scale, the star would move further inside the {\it lithium desert}.
Encouraged by this finding, we additionally searched the literature for new A(Li) determinations   for MS and subgiants stars from 2012 to date. In total, we found 17 works 
\citep{mishenina12,cumings12,li12,monaco12,pace12,schaeuble12,xing12,xing12b,boesgaard13,francois13,takeda13,gonzalez14,delgado14,
anna14,dobrovolskas14,anna14,gonzalez15}, whose results provide 2 more stars within the {\em lithium desert} boundaries: HD~44985 
\citep[$T_{\rm eff}$=6004~K and A(Li)=1.87~dex;][]{gonzalez14} and the star $\#$58440 of M4 \citep[$T_{\rm eff}$=5979~K and A(Li)=1.82~dex;][]{monaco12}.   
Keeping in mind that the A(Li) values come from different works, we can point out two possible conclusions regarding the existence of a {\em lithium desert}:
\begin{itemize}
\item[i)] If there actually is a short-lived mechanism responsible for the {\em lithium desert}, as suggested by \citet{ramirez12}, the stars BD+47~3218, HD~44985 and $\#$58440 in M4 represent a good benchmark to investigate it.
\item[ii)] The {\em lithium desert} is just a statistical product.
\end{itemize}

\subsection*{The Li-rich star BD+28~4515.} 
\begin{figure}
\includegraphics[width=75mm]{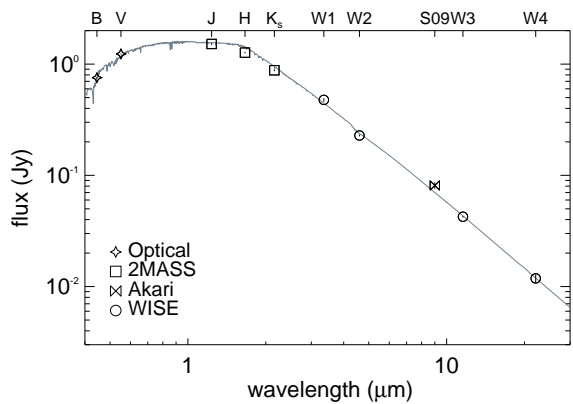}
\caption{Optical and infrared SED of BD+28~4515 . The symbols show the observed photometry, while the solid line represents the synthetic photospheric emission. Error bars are plotted, but they are in most cases much smaller than the symbol size.}
\label{fig:sed}
\end{figure}

The star with the highest lithium abundance in our sample is BD+28~4515, with A(Li)=3.05~dex. Comparing with the \cite{ramirez12} data, this value is 
the highest in a bin of temperature of 400~K, from 5350 to 5750~K. Within this $T_{\rm eff}$ range there is a total of 228 objects, from the \cite{ramirez12} and our samples, emphazising the peculiarity of BD+28~4515.
From our literature revision, we found 3 more stars in the same $T_{\rm eff}$ interval with A(Li)$>$3.0~dex: HD~287927 \citep[A(Li)=3.24~dex;][]{xing12}, BD+07~582 \citep[A(Li)=3.26~dex;][]{xing12}, and HD~285840 \citep[A(Li)=3.02~dex;][]{xing12b}. They are all classified as weak line T~Tauri stars \citep{xing10}.
Using the available data in the literature, we constructed the spectral energy distribution (SED) of BD+28~4515  and 
compared it, in Fig.~\ref{fig:sed}, with its synthetic SED interpolated in the \cite{castelli03} grid, adopting the stellar parameters reported in Table~\ref{tab:samp}. Even though we notice the lack of the distinctive infrared excess of a classical T-Tauri star, the AKARI point, which might reflect an emission of the 10~$\mu$m silicate feature, appears to be in excess, however at 2$\sigma$ significance only. We cannot rule out a possible weak line T-Tauri nature for BD+28~4515. Its high lithium abundance is compatible with the values of the Pleiades (see Fig.~2 of \citealt{deliyannis02}), which may be an indication of a relatively young age.

\section*{Acknowledgements}
We want to dedicate this paper to the memory of Don Hunten
who kindly donated the LPL Echelle Spectrograph to INAOE. We also thank
Ann Sprague who gave us a useful tutorial on the use of the
spectrograph. RLV, EB and MC would like to thank CONACyT for financial support through grants SEP-2011-169554 and SEP-2009-134985. This research has
made use of the SIMBAD data base, operated at CDS, Strasbourg, France.

\label{lastpage}

\end{document}